\documentclass[12pt,preprint,preprintnumbers,amsmath,amssymb,prb,superscriptaddress]{revtex4}
\usepackage{graphicx}

\newcommand{\qexp}{ \exp_{q} }
\newcommand{\abs}[1]{\lvert #1\rvert}
\newcommand{\re}{ \mathfrak{Re} }
\newcommand{\dt}{\Delta t}
\newcommand{\dx}{\Delta x}
\newcommand{\ket}[1]{|{#1}\rangle}

\begin{document}
\title{Discrete-time quantum walk with feed-forward quantum coin}
\author{Yutaka Shikano}
\email{yshikano@ims.ac.jp}
\affiliation{Research Center of Integrative Molecular Systems (CIMoS), Institute for Molecular Science, Okazaki, Aichi 444-8585, Japan}
\affiliation{Institute for Quantum Studies, Chapman University, Orange, California 92866, USA}

\author{Tatsuaki Wada}
\email{wada@mx.ibaraki.ac.jp} 
\affiliation{Department of Electrical and Electronic Engineering, Ibaraki University, Hitachi, Ibaraki
316-8511, Japan}

\author{Junsei Horikawa}
\email{12nd109l@hcs.ibaraki.ac.jp} 
\affiliation{Major in Materials Science, Graduate School of Science and Engineering, Ibaraki University, Hitachi, Ibaraki 316-8511, Japan}
\date{\today}

\begin{abstract}
Constructing a discrete model like a cellular automaton is a powerful method 
for understanding various dynamical systems.  However, the relationship between 
the discrete model and its continuous analogue is, in general, nontrivial. 
As a quantum-mechanical cellular automaton, a discrete-time quantum walk is defined 
to include various quantum dynamical behavior. Here we generalize a discrete-time 
quantum walk on a line into the feed-forward quantum coin model, which 
depends on the coin state of the previous step. We show that our proposed 
model has an anomalous slow diffusion characterized by the porous-medium equation, 
while the conventional discrete-time 
quantum walk model shows ballistic transport.
\end{abstract}
\maketitle

\section*{Introduction}
Cellular automata -- discrete models that follow a set of rules~\cite{Neumann} -- have 
been analyzed in various dynamical systems in physics, as well as in computational 
models and theoretical biology; well-known examples include crystal growth and the 
Belousov-Zhabotinsky reaction. To simulate quantum mechanical phenomena, 
Feynman~\cite{Feynman} proposed a quantum cellular automaton (the Feynman checkerboard).  
This model, defined in the general case by Meyer~\cite{Meyer}, is known as the discrete-time 
quantum walk (DTQW). Since the DTQW on a graph is a model of a universal quantum 
computation~\cite{Lovett, Childs2}, it is of great utility, especially in quantum 
information~\cite{kempe, salvador,kitagawa, shikano}. Furthermore, the DTQW has 
been demonstrated experimentally in various physical systems~\cite{Karski, Yaron, Photon,
Roos, Kitagawa, Photon0,Photon00,Photon01,Photon2, Photon3, Photon4, Photon5, Photon6, Fukuhara, PhysicalQW} to reveal 
quantum nature under dynamical systems. 

As the cellular automaton can be mapped to various differential equations by 
taking the continuous limit, some DTQW models can be mapped to the Dirac 
equation~\cite{Strauch, Sato, CBS}, the spatially discretized Schr\"{o}dinger 
equation~\cite{CKSS, Childs}, the Klein-Gordon equation~\cite{Debbasche, CBS}, or 
various other differential equations~\cite{Knight,Valcarcel}. These equations have 
ballistic transport properties, which are reflected mathematically in the 
one-dimensional (1D) DTQW with a time- and spatial-independent coin operator, 
i.e. a 1D {\it homogeneous} DTQW~\cite{Konno}. We consider here the 1D DTQW model. 
Physically, the standard deviation of the homogeneous DTQW is $\sigma
(t) \sim t$, whereas the unbiased classical random walk has a standard deviation of
$\sigma (t) \sim \sqrt{t}$.

In the homogeneous DTQW, the time evolution of a quantum particle
(walker) is given by a unitary operator $U$ defined on the composite
Hilbert space $\mathcal{H}_s \otimes \mathcal{H}_c$, where
$\mathcal{H}_s := {\rm span} \{ \ket{j}, j \in \mathbb{Z} \}$ is the
walker Hilbert space, and $\mathcal{H}_c$ is the two-dimensional
coin Hilbert space. For a unitary operator $U$, 
the quantum state evolves in each time step $t$
by
\begin{align}
  \ket{\Psi^{t+1}} = U \ket{\Psi^t}
\end{align}
with
\begin{align}
   \ket{\Psi^t} = \sum_{j=-\infty}^{\infty} \ket{j} \otimes
  \begin{pmatrix}
     a_j^t \\
     b_j^t
  \end{pmatrix},
\end{align}
where the upper $a_j^t$ (lower $b_j^t$) component corresponds to the
left (right) coin state at the $j$-th site at time step $t$. As an example, the
time evolution of the DTQW is given by
\begin{align}
  a_{j-1}^{t+1} &= \cos \theta \, a_j^t -\sin \theta \, b_j^t, \notag \\
  b_{j+1}^{t+1} &= \sin \theta  \, a_j^t + \cos \theta \, b_j^t.
\label{DTQW}
\end{align}
The $j$-th site probability at time step $t$ is given by
$P_j^t = |a_j^t|^2 + |b_j^t|^2$, and $\sum_{j=-\infty}^{\infty}
P_j^t = 1$ is satisfied for each time step $t$.

As a generalization of Eq.~\eqref{DTQW}, we define a DTQW with a 
feed-forward quantum coin described by
\begin{align}
  a_{j-1}^{t+1} &= g_j^t \, a_j^t -  \sqrt{1- \abs{ g_j^t }^2} \, b_j^t,
\notag
\\
  b_{j+1}^{t+1} &= \sqrt{1-\abs{g_j^t}^2} \, a_j^t + (g_j^t)^{\star} \, b_j^t,
 \label{nostro_modello}
\end{align}
with the site-dependent rate function
\begin{align}
  g_j^t = \abs{ a_{j-1}^t } + i \, \abs{ b_{j+1}^t },
\end{align}
which incorporates the nearest-neighbor interactions. 
Since this quantum coin depends on the probability distribution 
of the coin states on the nearest-neighbor sites 
at the previous step, this model is called a {\it feed-forward DTQW}. It is remarked that 
the feed-forward DTQW is one of the nonlinear DTQW models. Note that if
we set the rate function $g_j^t$ to $g = \cos \theta$, which is time
and site independent, then the model in Eq.~\eqref{nostro_modello}
reduces to the homogeneous model in Eq.~\eqref{DTQW}. We will 
show that our proposed feed-forward DTQW is experimentally feasible.
Furthermore, we will show that this model shows the anomalous diffusion 
as introduced below.

One of the famous anomalous diffusion equations is the 
porous medium equation (PME)~\cite{Vazquez}, defined by
\begin{align}
   \frac{\partial}{\partial t} p(x,t)
   = \frac{\partial^2}{\partial x^2} \;  p^{m}(x,t),
  \label{PME}
\end{align}
where the real parameter $m > 1$ characterizes the degree of porosity of
the porous medium. It is known that the PME can be derived from
three physical equations for the density $\rho$, pressure $p$, and
velocity ${\bf v}$ of the gas flow: the equation of continuity,
$\partial \rho / \partial t + \nabla \cdot (\rho {\bf v}) = 0$;
Darcy's law, ${\bf v} \propto - \nabla p$; and the equation of state
for a polytropic gas, $p \propto \rho^{\nu}$, where $\nu$ 
is the polytropic exponent and $m = \nu + 1$. One of the peculiar features of the
PME is the so-called \textit{finite propagation}, which implies the
appearance of a \textit{free boundary} separating the positive
region ($p > 0$) from the empty region ($p=0$). 

A well-known solution of the PME is the Barenblatt-Pattle (BP) one
\cite{Barenblatt}; it is self-similar, and its total mass is conserved during evolution. 
The evolutionary behavior of the BP
solution was recently studied in the context of generalized
entropies and information geometry~\cite{OW10}. The BP solution can
also be expressed by Tsallis' one-real-parameter ($q$)
generalization of a Gaussian function, i.e., the $q$-Gaussian~\cite{Tsallis}. In
the case of 1D space, the BP solution is
\begin{align}
 p_q(x,t)
  \propto \left[1-(1-q) \frac{x^2}{\sigma_q^2(t)} \right]^{\frac{1}{1-q}}
 \equiv \qexp\left( -\frac{x^2}{\sigma_q^2(t)} \right),
 \label{q-Gaussian}
\end{align}
with $q = 2-m$. Here, $\sigma_q^2(t)$ is a positive parameter that
characterizes the width of the $q$-Gaussian at time $t$ and is
similar to the variance $\sigma_{q=1}^2 (t)$ in a standard Gaussian. In
other words, the parameter $\sigma_q(t)$ characterizes the spread of
the $q$-Gaussian distribution \cite{Anteneodo,SNT08};
\begin{align}
   \sigma_q (t) \propto t^{\frac{1}{3-q}},
  \label{sig_q4PME}
\end{align}
which reduces to $\sigma_{q=1} (t) \propto \sqrt{t}$ in the limit of
$q \to 1$. Note that in the same limit, the $q$-Gaussian reduces to
the standard Gaussian, $\exp\left( -x^2 / \sigma_{q = 1}^2(t) \right)$,
and the PME reduces to the standard heat equation $\partial p /
\partial t = \partial^2 p / \partial x^2$.

In this paper, we analyze a specific feed-forward DTQW with an 
experimental proposal using the polarized state and optical mode. We show
numerically that the probability distributions of the feed-forward DTQW
model have anomalous diffusion characterized by $\sigma_{q=0.5} (t) \sim
t^{0.4}$. These dynamics are consistent with the time evolution of
the self-similar solution~\cite{Barenblatt} of the PME, which is 
known to describe well the anomalous diffusion of an
isotropic gas through a porous medium. Furthermore, we show
analytically that the interference terms in our model help 
the speedup of the associated Markovian model but does not help
the quadratic speedup like
the homogeneous DTQW does~\cite{Ro10}.
Note that although anomalous diffusion was found numerically in
a nonlinear model~\cite{NPR}, an aperiodic time-dependent coin  
model~\cite{RMM04}, and the history-dependent coin~\cite{Rohde} 
from the time dependence of the variance $\sigma_{q=1} (t)$, 
the partial differential equation (PDE) corresponding to 
their models have not derived due to the lack of the numerical step 
(about $100$ step). Therefore, we have not yet revealed the origin 
of the anomalous diffusion in the DTQW. 

\section*{Results}
\subsection*{Experimental proposal of feed-forward DTQW}
We propose an optical implementation of the feed-forward DTQW. In the simple 
optical implementation of the homogeneous DTQW, the walker space uses the 
spatial mode and the coin space does the polarized state. The shift uses the 
polarized beam splitter and the quantum coin uses the quarter-wave, half-wave, 
and quarter-wave plates, which can arbitrarily 
rotate the polarized state in the Poincar\'e sphere. This was 
experimentally done in Refs.~\cite{Photon0,Photon00,Photon01,Photon, Kitagawa, Photon2, Photon3, Photon4, Photon5, Photon6}.

\begin{figure}[ht]
\includegraphics[width=150mm]{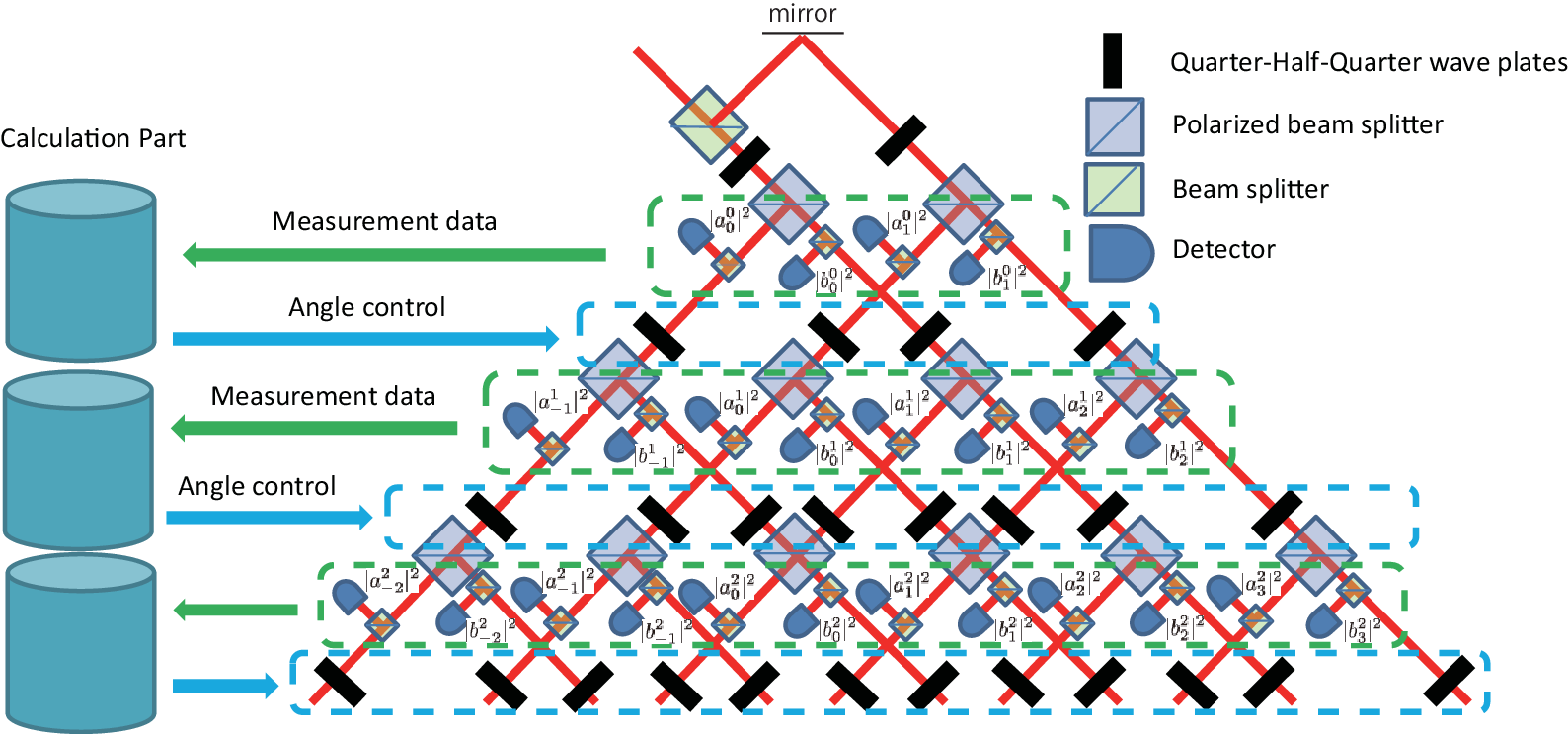}
\caption{\label{fig:opt} {\bf Optical implementation of the feed-forward DTQW model.} 
Figure shows our experimental proposal of 
our model. From the intensity of the detectors for each path, the polarizers should be changed. This 
can be taken as the feed-forward quantum coin.}
\end{figure}

Let us construct the feed-forward system of the quantum coin.
The detectors put at each path to evaluate the probability distribution of the coin 
state $|a^t_j|^2$ and $|b^t_j|^2$. Since our proposed quantum coin depends on $|a^t_j|$ 
and $|b^t_j|$, we can calculate the coin operator at the $j$th site. According to 
the Jones calculation~\cite{Yariv} to satisfy Eq. (\ref{nostro_modello}), we control the angels of the 
quarter-wave, half-wave, and quarter-wave plates for each path. This can be taken as the quantum 
coin operator with the feed-forward. This is depicted in Fig.~\ref{fig:opt}. 
In what follows, we consider the long time time evolution of the feed-forward DTQW.
\subsection*{Numerical results of feed-forward DTQW with anomalous diffusion}
To study the time evolution of the feed-forward DTQW model, the initial
state should have nonzero coin states at the nearest-neighbor
sites. This can be easily understood by considering the following
example. Let us take $(a_0^0, b_0^0)$ as the only non-zero initial
state. In this case, the rate is $g_0^0 = 0$, because there is no
neighboring state. From the map in Eq.~\eqref{nostro_modello}, we
see that the nonzero states at $t=1$ are $a_{-1}^{1} = -b_0^0$ and
$b_{1}^{1} = a_0^0$. This gives $g_{-1}^1 = g_1^1=0$, and we see
that the only nonzero state is $(a_0^2, b_0^2) = (-b_1^1, a_{-1}^1)
= (-a_0^0, -b_0^0)$ at $t=2$. This state at $t=2$ only differs in
sign (or phase) from the initial state. Thus if the initial state is
concentrated at a single site, no spreading occurs; the state
only oscillates around the initial site.

\begin{figure}[ht]
\includegraphics[width=150mm]{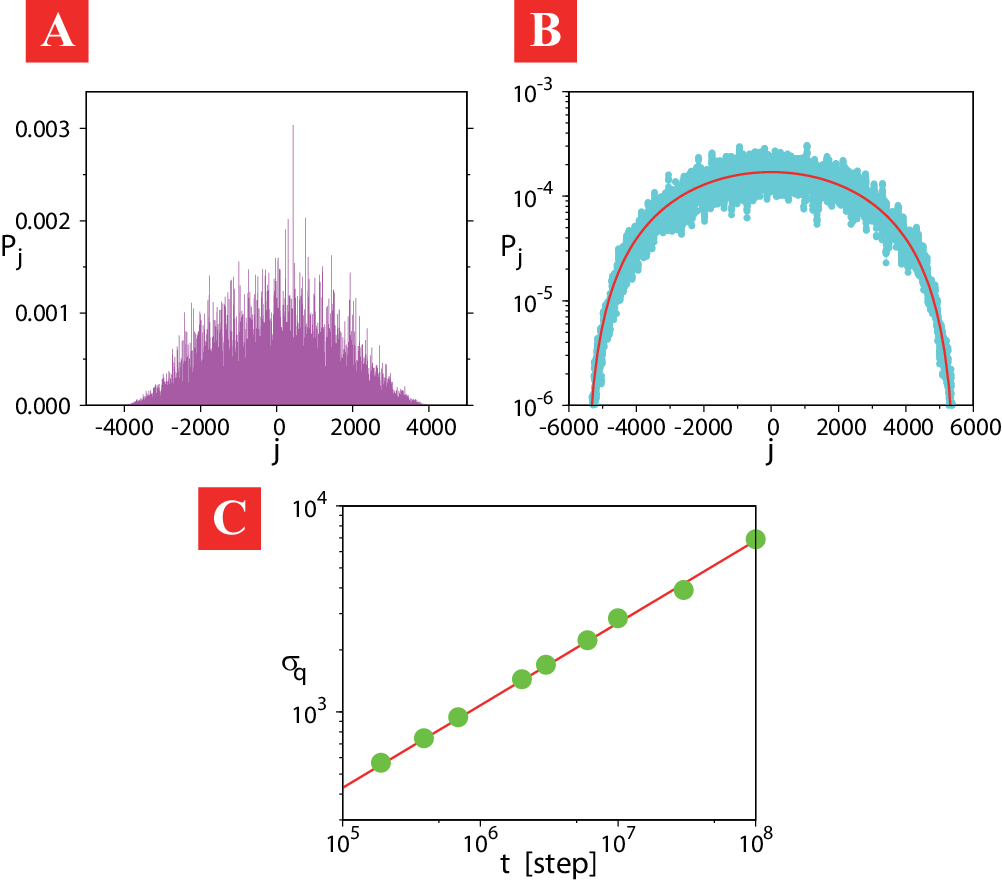}
\caption{\label{fig:data2} {\bf Anomalous slow diffusion of the
feed-forward DTQW model.} Its probability distribution at $t=10^7$ step displayed in Panel (A) 
with running averaged over 10 data sets (light blue line) is fitted by the
$q$-Gaussian \eqref{q-Gaussian} with $q=0.5$ (red line) to obtain the
$q$-generalized standard deviation $\sigma_q(t)$ in Panel (B). Panel (C) shows 
the long-time evolution of the $q$-generalized standard deviation 
$\sigma_q(t)$ (green dots), which is well fitted by $\sigma_{q=0.5}(t) \sim t^{0.4} $ 
(red line).}
\end{figure}
Figure~\ref{fig:data2} (A) shows a typical probability distribution of
the feed-forward DTQW after a long-time evolution. See the Supplementary 
Movie for more details. The initial state
was set as $(a_0^0, b_0^0) = (a_1^0,b_1^0) = (1/2, i/2)$.
We note that the probability distribution diffuses very slowly and
does not approach a Gaussian. These features are often observed in
anomalous diffusion. It is also remarked that such behavior has not 
yet seen in DTQWs with the position-dependent coin~\cite{Ro09, Mc, Joye, SK}, 
which show the localization property.

We performed long-time numerical simulations of the feed-forward DTQW
model [Eq.~\eqref{nostro_modello}] for up to $t \sim 10^8$ steps. To
study the asymptotic behavior, we take running averages of the
numerical solutions to reduce the influence of multiple spikes. The
averaged data were fitted with the $q$-Gaussian of
Eq.~\eqref{q-Gaussian} to determine the corresponding
$q$-generalized standard deviation $\sigma_q(t)$, as shown in
Fig.~\ref{fig:data2} (B). We note that the averaged data at each time
step are well fitted by the $q$-Gaussian with $q=0.5$.

The long-time evolution of $\sigma_q(t)$, plotted in
Fig.~\ref{fig:data2} (C), reveals that the time evolution of the
feed-forward DTQW model is well characterized by $\sigma_{q=0.5}(t)
\sim t^{0.4}$, which is the same time dependency for $q=0.5$ of the
PME [Eq.~\eqref{sig_q4PME}].

\subsection*{Analytical derivation of anomalous diffusion in the associated Markov model of feed-forward DTQW}
The relationship between our model and the PME can be explored using
the decomposition method of Romanelli {\it et al.}~\cite{Ro04,Ro10}, in
which the unitary evolution of a DTQW model is decomposed into
Markovian and interference terms. We obtain the following map for
both coin distributions $\abs{ a_{j}^{t}}^2$ and $\abs{ b_{j}^{t}}^2
$:
\begin{align}
  \abs{ a_{j-1}^{t+1}}^2 &= \abs{g_j^t}^2 \, \abs{a_j^t}^2
   +  (1- \abs{ g_j^t }^2) \, \abs{b_j^t}^2
   - 2 \sqrt{1-\abs{g_j^t}^2} \; \beta_j^t,
\notag
\\
  \abs{ b_{j+1}^{t+1}}^2 &= (1-\abs{g_j^t}^2) \, \abs{a_j^t}^2
   +  \abs{ g_j^t }^2 \, \abs{b_j^t}^2
   + 2 \sqrt{1-\abs{g_j^t}^2} \; \beta_j^t,
\label{ab}
\end{align}
where the two terms including $\beta_j^t = \re[ g_j^t a_j^t
(b_j^t)^{\star}]$ are interference terms, and $\re[z]$ is the real
part of a complex number $z$.

\begin{figure}[ht]
\includegraphics[width=140mm]{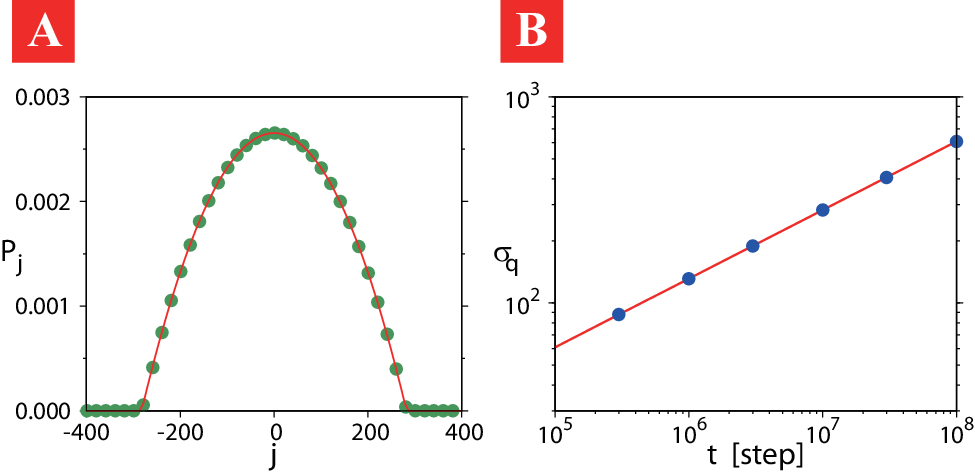}
\caption{\label{fig:Markov} {\bf Anomalous slow diffusion of the associated Markovian model for  
the nonlinear quantum walk.} 
Panel (A) shows the probability distribution of the associated Markovian model at 
$t=10^7$ step (green dots) fitted by the $q$-Gaussian, yielding 
$q=0.0$ and $\sigma_{q = 0.0} (t)=283$ (red line). Panel (B) shows 
the long-time evolution of the $q$-generalized standard 
deviation $\sigma_q(t)$ of the associated Markovian model (blue dots). 
It is well fitted by $\sigma_{q=0.0}(t) \sim t^{0.33} $ (red line).}
\end{figure}

Neglecting the interference terms and introducing the abbreviations
$L_j^t = \abs{ a_j^{t}}^2$ and $R_j^t = \abs{ b_j^{t}}^2$, we get the
{\it associated Markovian model};
\begin{align}
 R_{j+1}^{t+1} + L_{j-1}^{t+1} &= R_j^t + L_j^t,
\label{a'}
\\
  R_{j+1}^{t+1} - L_{j-1}^{t+1} &=
\left\{ 2 (L_{j-1}^t + R_{j+1}^t) -1 \right\} (R_j^t - L_j^t ).
\label{b'}
\end{align}
The numerical simulation of the associated Markovian model is
performed under initial conditions of $ (R_0^0, L_0^0) = (R_1^0,
L_1^0) = 1/4$, and the typical probability distribution shown in
Fig.~\ref{fig:Markov} (A) is well fitted by the $q$-Gaussian with $q=0.0$.
Furthermore, Fig.~\ref{fig:Markov} (B) shows that the time evolution of
$\sigma_q(t)$ of the associated Markovian model is well fitted to
$\sigma_{q=0.0}(t) \sim t^{0.33} $, which again is the same time
dependency as the PME for $q=0$.

It is known that the classical Markovian model, i.e. one without
the interference terms of the homogeneous DTQW, satisfies the
standard heat equation in the continuous limit. Consequently, the
associated asymptotic probability distribution is a standard
Gaussian. This implies that the ballistic transport property of the
homogeneous DTQW comes from the interference term~\cite{Ro10}. We thus consider
the continuous limit~\cite{GG96} of the associated Markovian model.

We introduce the density $\rho(x,t)$ and current $j(x,t)$ as
\begin{align}
  \rho(x,t) = L_j^t + R_j^t, \quad j(x,t) = (R_j^t - L_j^t)/ \dx,
\end{align}
where $\dx$ is the difference of the nearest-neighbor sites. Taking
a Taylor expansion of Eq.~\eqref{a'}, we get
\begin{align}
 \frac{\partial}{\partial t} \rho(x,t) + \frac{\partial}{\partial x} j(x,t)
  + \frac{1}{2} \frac{\partial^2}{\partial x^2} \rho(x,t) = 0,
  \label{fluid-eq}
\end{align}
in the diffusion limit, i.e., the quantity $(\dx)^2 / \dt$ remains
constant (set to unity here for simplicity) as $\dt, \dx \to 0$ with
the one-step time difference $\dt$. In a similar manner, by
expanding Eq.~\eqref{b'} and taking the diffusion limit, we obtain
\begin{align}
 j(x,t)
  = -\frac{1}{2 \big(1-\rho(x,t)\big)}\frac{\partial}{\partial x} \rho(x,t),
 \label{j}
\end{align}
which implies a breakdown in Fick's first law ($j \propto -\partial
\rho / \partial x$) and is the hallmark of anomalous diffusion. By
substituting Eq.~\eqref{j} into Eq.~\eqref{fluid-eq}, we obtain the
following nonlinear PDE:
\begin{align}
 \frac{\partial}{\partial t} \rho(x,t) &=
\frac{1}{2 \big( 1-\rho(x,t) \big)^2} \notag \\
   \times & \left( \frac{1}{2} \frac{\partial^2}{\partial x^2} \rho^2(x,t)
- \rho^2(x,t) \frac{\partial^2}{\partial x^2} \rho(x,t) \right).
\label{nlpde}
\end{align}
Evaluating the asymptotic solution of this nonlinear PDE, after a
long-time evolution, $\rho(x,t)$ becomes much less than unity. As the rough 
approximation in this long-time limit, we have $1-\rho \approx 1$ and $\rho^2 \approx
0$, and Eq.~\eqref{nlpde} is thus well approximated by
\begin{align}
 \frac{\partial}{\partial t} \rho(x,t) \approx
     \frac{1}{4} \frac{\partial^2}{\partial x^2} \rho^2(x,t),
\label{approx}
\end{align}
which is nothing but the PME in Eq.~\eqref{PME} with $m=2$ ($q=0$).
We thus conclude that the approximated asymptotic solution of
Eq.~\eqref{nlpde} is a $q$-Gaussian with $q=0$. In addition, we can
show that this result is mathematically valid by applying the asymptotic 
Lie symmetry method~\cite{Gaeta} (see Method). This method can give 
an equivalence between the asymptotic solution of the PDE and the 
analytically-solved one of the other PDE without analytically 
solving this PDE. Therefore, the associated Markovian model exhibits 
anomalous diffusion
described by the PME in Eq.~\eqref{PME} with $m=2$. This implies 
that the interference term of our model leads to 
the speed-up of the quantum walker $\sigma_{q=0.5} \sim t^{0.4}$ 
compared to the associated Markovian model $\sigma_{q=0} 
\sim t^{1/3}$ and makes the zig-zag shape around the $q$-Gaussian
distribution.

In summary, we have proposed a feed-forward DTQW model
Eq.~\eqref{nostro_modello} in which the coin operator depends on the
coin states of the nearest-neighbor sites. We show that this model is 
experimentally feasible. Our feed-forward DTQW model
asymptotically satisfies the PME for $m = 1.5 \, (q = 0.5)$ and
exhibits anomalous slow diffusion $\sigma_{q=0.5} (t) \sim t^{0.4}$ from
the probability distribution and the time dependency of the standard deviation 
defined in the $q$-Gaussian distribution. 

\section*{Discussion}
\begin{figure}[ht]
\begin{center}
\includegraphics[width=130mm]{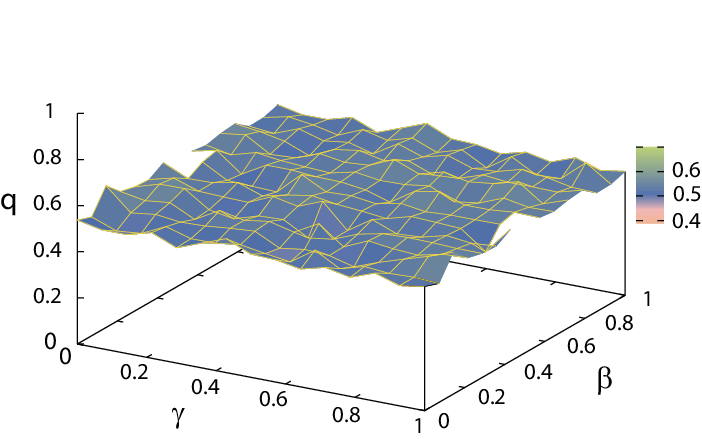}
\caption{\label{fig:initial} {\bf Initial coin state dependence.} 
Changing the parameters $\beta$ and $\gamma$, we numerically evaluate the parameter $q$ of $q$-Gaussian distribution for $225$ different initial states expressed by $(a_0^0, b_0^0) = (\cos \beta \pi / \sqrt{2}, \sin \beta \pi / \sqrt{2})$ and $(a_1^0,b_1^0) = (\cos \gamma \pi / \sqrt{2}, \sin \gamma \pi / \sqrt{2})$. Note that the trivial cases, $\beta = 0.5, \gamma = 0$ and $\beta = 0.5, \gamma = 1$, are not plotted. Our fitting result except for the trivial cases is $q= 0.5^{+0.116}_{-0.047}$.}
\end{center}
\end{figure}

In this section, we show that our results after the long-time numerical simulations 
have no initial coin dependence, and 
that the interference term can be taken as the noise source in addition to the PME. 
First, while the above analysis uses the only fixed initial coin states as $(a_0^0, b_0^0) = (a_1^0,b_1^0) = (1/2, i/2)$, 
we numerically confirm that there is almost no dependence of the initial coin state except for the trivial cases as follows. 
We have performed the several numerical simulations for the initial state specified by
$(a_0^0, b_0^0) = (\cos \beta \pi / \sqrt{2}, \sin \beta \pi / \sqrt{2})$ and 
$(a_1^0,b_1^0) = (\cos \gamma \pi / \sqrt{2}, \sin \gamma \pi / \sqrt{2})$ with the real-parameter $\beta$ and $\gamma$ ranging from
$0$ to $1$. 
Note that the trivial cases, $\beta = 0.5, \gamma = 0$ and $\beta = 0, \gamma = 0.5$, lead to the localization of the probability distribution for any time, 
and we cannot define the parameter $q$ for the trivial initial states. 
Figure~\ref{fig:initial} shows the numerical evaluation of the parameter $q$ of  
$q$-Gaussian distribution from the data at the two different time steps $t = 10^6$ and $t=10^7$, 
under the assumption to satisfy the stationary solution 
of the PME [Eqs. (\ref{q-Gaussian}) and (\ref{sig_q4PME})].
The evaluated $q$-parameters for the various initial states are $q= 0.5^{+0.116}_{-0.047}$ except for the trivial cases.
Therefore, we can conclude that our nonlinear model shows the anomalous slow diffusion to satisfy the PME 
with $m \simeq 1.5 \, (q \simeq 0.5)$ without the initial state dependence. 

\begin{figure}[ht]
\begin{center}
\includegraphics[width=100mm]{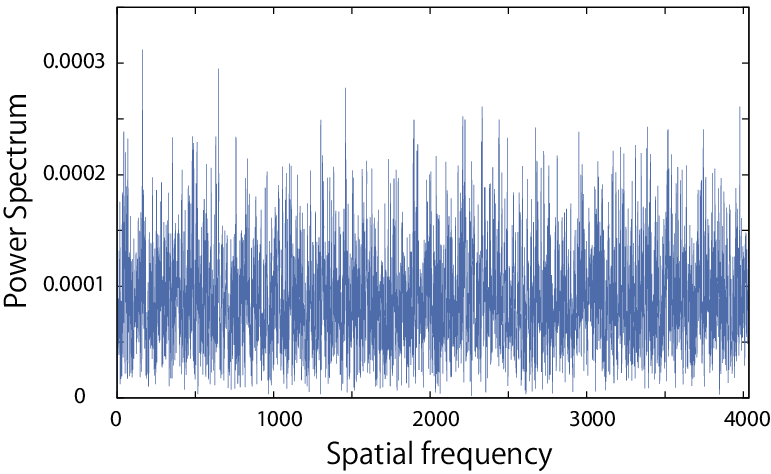}
\caption{\label{fig:power} {\bf The difference between the nonlinear model and the fit.} 
The power spectrum of the difference between the probability distribution of our model and
the $q$-Gaussian with $q=0.5$ at $10^7$ step. To remove the effects
of the expectation value, we replace $x$ with $x - 36.91$ in 
the $q$-Gaussian with $q = 0.5$ [Eq. \eqref{q-Gaussian}].}
\end{center}
\end{figure}

Finally, let us consider the difference between the probability distribution of our
model and the $q$-Gaussian distribution with $q=0.5$, as shown in
Fig.~\ref{fig:data2} (B); the power spectrum of this difference exhibits a white noise 
as shown in Fig.~\ref{fig:power}. This power spectrum divided by the
physical time scale $t^{0.4}$ may remain finite in the asymptotic
case, which suggests that our nonlinear model may be mapped to the
stochastic PME, i.e. the PME plus a white noise term, in the
continuous limit. This stochasticity must come from the interference
term. The problem of extracting the stochasticity from a
deterministic process has been discussed in another context, that of
Mori's noise~\cite{mori}. Further analysis of this model may
reveal the origin of the stochasticity. This is interesting as a purely mathematical 
problem of a stochastic nonlinear partial differential equation and for showing 
the relationship between the discrete model and its continuous limit.

\section*{Method}
In what follows, the solution of Eq. (\ref{nlpde}) is asymptotically identical to 
the solution of Eq. (\ref{approx}). This is mathematically equivalent to 
showing that the probability distribution
\begin{align}
  \rho^{(q = 0)} (x)= \frac{1}{Z(\sigma_{q = 0})} \left[1 - \frac{x^2}{\sigma_{q = 0}^2} \right],
 \label{rho0}
\end{align}
is invariant under an asymptotic Lie symmetry \cite{Gaeta} of the
nonlinear partial differential equation (\ref{nlpde}). In other words,
\begin{align}
    \partial_{t} \rho =
    \frac{1}{2(1-\rho )^2} \left[ \frac{1}{2} \partial_{xx} \rho^2 - \rho^2 \partial_{xx} \rho \right].
    \label{ourPDE}
\end{align}
In Eq.~\eqref{rho0}, $Z(\sigma_{q =0}) = 4 \sigma_{q = 0} / 3$ is
the normalization factor, and in what follows, the argument of this
function is omitted where possible and $\partial_{t} \rho$ is
denoted as $\rho_t$ for simplicity.

We follow the asymptotic Lie symmetry method and notations in Ref.
\cite{Gaeta}. Under an infinitesimal transformation with the
generator
\begin{equation}
X = \xi \, \partial_x + \tau \, \partial_t + \varphi \, \partial_{\rho},
\end{equation}
that is
\begin{align}
  x &\mapsto \hat{x} = x + \epsilon \, \xi(x, t, \rho), \notag \\
  t &\mapsto \hat{t} = t + \epsilon \, \tau(x, t, \rho), \notag \\
 \rho &\mapsto \hat{\rho} = \rho + \epsilon \, \varphi(x, t, \rho),
\end{align}
the function $\rho(x,t)$ is mapped to a new function
$\hat{\rho}(x,t)$, with
\begin{equation}
 \hat{\rho}(x,t) = \rho(x,t)
+ \epsilon \, \big[\varphi - \rho_x \xi - \rho_t \tau \big]_{\rho=\rho(x,t)}.
\end{equation}
By applying this to the probability distribution Eq.~\eqref{rho0},
we see that the transformation $X$ with $\xi = -x$ leaves
Eq.~\eqref{rho0} invariant if and only if
\begin{equation}
\varphi = \rho_x \xi = - \rho_x x =  \frac{2 x^2}{ Z \sigma_{q=0}^2}.
\end{equation}
Note that $\tau=\eta \cdot t$ remains unrestricted at this stage
because $\rho^{(q = 0)}(x)$ does not explicitly depend on time $t$.
Conversely, the function $\rho(x)$ is invariant under $X = -x
\partial_x + \tau \partial_t + 2 x^2 /(Z \sigma_{q=0}^2) \, \partial_{\rho}$ for any $\tau$ 
if and only if $\rho(x)$ is of the form given in Eq.~\eqref{rho0}.

Following the general procedure for a Lie group analysis of
differential equations~\cite{Olver}, the second prolongation of $X$
is described by
\begin{equation}
   Y = X + \Psi_t \partial_{\rho_t} + \Psi_x \partial_{\rho_x} + \Psi_{xx} \partial_{\rho_{xx}}.
\end{equation}
The coefficients $\Psi_t, \Psi_x$, and $\Psi_{xx}$ are defined as
follows. Under an infinitesimal transformation of $X$, the partial
derivatives are transformed as $\rho_x \mapsto \rho_x + \epsilon \,
\Psi_x$, $\rho_t \mapsto \rho_t + \epsilon \, \Psi_t$, and
$\rho_{xx} \mapsto \rho_{xx} + \epsilon \, \Psi_{xx}$. We then
readily obtain
\begin{equation}
  \varphi_x = \frac{4 x}{Z \sigma_{q=0}^2}, \quad
  \varphi_{xx} = \frac{4}{Z \sigma_{q=0}^2}, \quad
  \varphi_\rho =0, \quad
  \varphi_{\rho\rho} =0.
\end{equation}
The coefficients $\Psi^t, \Psi^x$, and $\Psi^{xx}$ are then obtained
by applying the prolongation formula (2.39) from Ref. \cite{Olver}:
\begin{align}
  \Psi^t &= (\varphi_\rho - \tau_t) \rho_t = - \eta \rho_t,\\
  \Psi^x &= \varphi_p + (\varphi_\rho -\xi_p) \rho_p
  = \frac{4x}{Z \sigma_{q=0}^2} + \rho_x, \\
  \Psi^{xx} &= \varphi_{xx} + 2 \varphi_{x\rho} \rho_x +
  \varphi_{\rho\rho} \rho_x^2 + (\varphi_\rho - 2 \xi_x) \rho_{xx} = \frac{4}{Z \sigma_{q=0}^2} + 2 \rho_{xx}.
\label{psi_pp}
\end{align}

We note that Eq.~\eqref{ourPDE} can be written as
\begin{equation}
  \rho_t = C_1 \, (\rho_x)^2 + C_2 \, \rho_{xx}
\end{equation}
with
\begin{equation}
    C_1 =  \frac{1}{2(1-\rho)^2}, \quad
    C_2 = \frac{\rho}{2(1-\rho)}.
\end{equation}
The asymptotic Lie symmetry condition
\begin{equation}
  Y \big( \rho_t -C_1 (\rho_x)^2 -C_2 \rho_{xx} \big) =
   \Psi^t - 2 C_1 \rho_x \Psi^x  -C_2 \Psi^{xx}
   - C_1' \varphi (\rho_x)^2 - C_2' \varphi \rho_{xx} = 0
 \label{condALS}
\end{equation}
with
\begin{equation}
    C_1' = \partial_\rho C_1 =  \frac{1}{2 (1-\rho)^3}, \quad
    C_2' = \partial_\rho C_2 =  \frac{1}{2(1-\rho)^2},
\end{equation}
can be written in the following compact form:
\begin{equation}
A_0(x,t,\rho) + A_1(x,t,\rho) \rho_x + A_2(x,t,\rho) (\rho_x)^2 + A_3(x,t,\rho) \rho_{xx} = 0.
\end{equation}
When the condition in Eq.~\eqref{condALS} is fulfilled, each $A_k
(k=0,1,2,3)$ function must vanish separately in the asymptotic limit
\begin{equation}
   \abs{\rho(x,t)} \to 0 \quad \textrm{for} \quad \abs{x} \to \infty,
\end{equation}
implying that the variance $\sigma_{q=0}$ also becomes infinity in
the asymptotic limit from Eq.~\eqref{rho0};
\begin{equation}
   \sigma_{q=0} \to \infty \quad \textrm{for} \quad \abs{x} \to \infty.
\end{equation}

The function $A_3$ can be expressed as
\begin{equation}
A_3  = \frac{1}{2 (1-\rho)} \left\{ \rho \cdot (\eta + 2) +
\frac{4}{ (1-\rho) Z \sigma_{q=0}^2} \right\},
\end{equation}
which must be nonzero as $\sigma_{q=0} \to \infty$, unless we choose
\begin{equation}
   \eta = -2.
\end{equation}
Making this choice, $X$ becomes
\begin{equation}
X = \xi \, \partial_x -2 t \, \partial_t + \frac{2 x^2}{ Z \sigma_{q=0}^2}\,
\partial_{\rho},
 \label{choisedX}
\end{equation}
and $A_3$ reduces to
\begin{equation}
  A_3  = \frac{2}{(1-\rho)^2 Z \sigma_{q=0}^2}.
\end{equation}
Thus, $A_3 \to 0$ as $\sigma_{q=0} \to \infty$.

In a similar manner, $A_0, A_1$, and $A_2$ are given by
\begin{equation}
A_0  = \frac{2 \rho}{(1-\rho)Z \sigma_{q=0}^2},  \quad
A_1  = \frac{4}{(1-\rho)^2 Z \sigma_{q=0}^2},  \quad
A_2  = \frac{2}{(1-\rho)^3 Z \sigma_{q=0}^2},
\end{equation}
and all become zero as $\sigma_{q=0} \to \infty$. Therefore, we
conclude that the distribution in Eq.~\eqref{rho0} is an invariant
solution for the transformation $X$ of Eq.~\eqref{choisedX}, which
is an asymptotic symmetry for large $\vert x \vert$ of the nonlinear
partial differential equation Eq.~\eqref{ourPDE}.

\section*{Acknowledgments}
Y.S. thanks Masao Hirokawa for valuable discussions. This work was
partially supported by the Joint Studies Program of the
Institute for Molecular Science.

\begin{thebibliography}{99}
\bibitem{Neumann}
von Neumann, J.
 The general and logical theory of automata,
in {\it Cerebral Mechanisms in Behavior: The Hixon Symposium},
L. A. Jeffress (Ed.),  pp. 1--41 (John Wiley and Sons, New York, NY, 1951).

\bibitem{Feynman}
Feynman, R. P.
Space-time approach to non-relativistic quantum mechanics
{\it Rev. Mod. Phys.} {\bf 20}, 367--387 (1948).

\bibitem{Meyer}
Meyer, D.
From quantum cellular automata to quantum lattice gases
{\it J. Stat. Phys.} \textbf{85}, 551--574 (1996).

\bibitem{Lovett}
Lovett, N. B., et al.
Universal quantum computation using the discrete-time quantum walk,
{\it Phys. Rev. A} {\bf 81}, 042330 (2010).

\bibitem{Childs2}
Childs, A., Gosset, D., \& Webb, Z.
Universal Computation by Multiparticle Quantum Walk, 
{\it Science} {\bf 339}, 791--794 (2013).

\bibitem{kempe}
Kempe, J.
Quantum random walks - an introductory overview, 
{\it Contemp. Phys.} {\bf 44}, 307--327 (2003).

\bibitem{salvador}
Venegas-Andraca, S. E. 
Quantum walks: a comprehensive review,
{\it Quant. Inf. Proc.} {\bf 11}, 1015--1106 (2012).

\bibitem{kitagawa}
Kitagawa, T. 
Topological phenomena in quantum walks: elementary introduction to the physics of topological phases,
{\it Quant. Inf. Proc.} {\bf 11}, 1107--1148 (2012).

\bibitem{shikano}
Shikano, Y. 
From Discrete Time Quantum Walk to Continuous Time Quantum Walk in Limit Distribution,
{\it J. Comput. Theor. Nanosci.} {\bf 10}, 1558--1570 (2013).

\bibitem{Photon0}
Do, B., et al. 
Experimental realization of a quantum quincunx by use of linear optical elements, 
{\it J. Opt. Soc. Am. B} {\bf 22}, 499--504 (2005).

\bibitem{Photon00}
Zhang, P., et al.
Demonstration of one-dimensional quantum random walks using orbital angular momentum of photons,
{\it Phys. Rev. A} {\bf 75}, 052310 (2007).

\bibitem{Photon01}
Perets, H. B., et al.
Realization of Quantum Walks with Negligible Decoherence in Waveguide Lattices,
{\it Phys. Rev. Lett.} {\bf 100}, 170506 (2008).

\bibitem{Karski}
Karski, M., et al. 
Quantum Walk in Position Space with Single Optically Trapped Atoms, 
{\it Science} {\bf 325}, 174--177 (2009).

\bibitem{Yaron}
Peruzzo, A., et al. 
Quantum walks of correlated particles,
{\it Science} {\bf 329}, 1500--1503 (2010).

\bibitem{Roos}
Z\"{a}hringer, F., et al. 
Realization of a Quantum Walk with One and Two Trapped Ions,
{\it Phys. Rev. Lett.} {\bf 104}, 100503 (2010).

\bibitem{Photon}
Schreiber, A., et al. 
Photons Walking the Line: A Quantum Walk with Adjustable Coin Operations, 
{\it Phys. Rev. Lett.} {\bf 104}, 050502 (2010).

\bibitem{Kitagawa}
Kitagawa, T., et al. 
Observation of topologically protected bound states in photonic quantum walks, 
{\it Nat. Comm.} {\bf 3}, 882 (2012).

\bibitem{Photon2}
Schreiber, A., et al. 
A 2D Quantum Walk Simulation of Two-Particle Dynamics,
{\it Science} {\bf 336}, 55--58 (2012).

\bibitem{Photon5}
Sansoni, L., et al.
Two-Particle Bosonic-Fermionic Quantum Walk via Integrated Photonics,
{\it Phys. Rev. Lett.} {\bf 108}, 010502 (2012).

\bibitem{Photon3}
Crespi, A., et al. 
Anderson localization of entangled photons in an integrated quantum walk,
{\it Nat. Photon.} {\bf 7} 322--328 (2013).

\bibitem{Photon4}
Jeong, Y.-C., et al. 
Experimental realization of a delayed-choice quantum walk,
{\it Nat. Comm.} {\bf 4}, 2471 (2013).

\bibitem{Photon6}
Xue, P., et al.
Observation of quasiperiodic dynamics in a one-dimensional quantum walk of single photons in space,
e-print: arXiv:1312.0123 (2013).

\bibitem{Fukuhara}
Fukuhara, T., et al. 
Microscopic observation of magnon bound states and their dynamics. 
{\it Nature} {\bf 502}, 76-79 (2013).

\bibitem{PhysicalQW}
Manouchehri, K. \& Wang, J. 
{\it Physical Implementation of Quantum Walks}
(Springer, Berlin, 2014).
 
\bibitem{Strauch}
Strauch, F. W. 
Relativistic effects and rigorous limits for discrete- and continuous-time quantum walks,
{\it J. Math. Phys.} {\bf 48}, 082102 (2007).

\bibitem{Sato}
Sato, F. \& Katori, M. 
Dirac equation with an ultraviolet cutoff and a quantum walk,
{\it Phys. Rev. A} {\bf 81}, 012314 (2010).

\bibitem{CBS}
Chandrashekar, C. M., Banerjee, S. \& Srikanth, R. 
Relationship between quantum walks and relativistic quantum mechanics,
{\it Phys. Rev. A} {\bf 81}, 062340 (2010).

\bibitem{CKSS}
Chisaki, K., Konno, N., Segawa, E., \& Shikano, Y. 
Crossovers induced by discrete-time quantum walks,
{\it Quant. Inf. Comp.} {\bf 11}, 741--760 (2011).

\bibitem{Childs}
Childs, A. M. 
On the Relationship Between Continuous- and Discrete-Time Quantum Walk,
{\it Comm. Math. Phys.} {\bf 294}, 581--603 (2010).

\bibitem{Debbasche}
di Molfetta, G. \& Debbasch, F. 
Discrete time Quantum Walks: continuous limit and symmetries,
{\it J. Math. Phys.} {\bf 53}, 123302 (2012).

\bibitem{Knight}
Knight, P., Rold\'{a}n, E. \& Sipe, J. E. 
Propagating Quantum Walks: the origin of interference structures,
{\it J. Mod. Opt.} {\bf 51}, 1761--1777 (2004).

\bibitem{Valcarcel}
de Valcarc\'{e}l, G. J., Rold\'{a}n, E., \& Romanelli, A. 
Tailoring discrete quantum walk dynamics via extended initial conditions,
{\it New J. Phys.} {\bf 12}, 123022 (2010).

\bibitem{Konno}
Konno, N. 
Quantum Random Walks in One Dimension,
{\it Quant. Inf. Proc.} {\bf 1}, 345--354 (2002);
A new type of limit theorems for the one-dimensional quantum random walk, 
{\it J. Math. Soc. Jpn.} {\bf 57},  935--1234 (2005).

\bibitem{Vazquez}
Vazquez, J. L. 
\textit{The Porous Medium Equation, Mathematical Theory}
(Oxford University Press, Oxford, 2006).

\bibitem{Barenblatt}
Barenblatt, G. I. 
\textit{Scaling, Self-Similarity, and Intermediate
Asymptotics} (Cambridge University Press, Cambridge, 1996).

\bibitem{OW10}
Ohara, A. \& Wada, T. 
Information geometry of q-Gaussian densities and behaviors of solutions to related diffusion equations,
{\it J. Phys. A} \textbf{43} 035002 (2010).

\bibitem {Tsallis}
Tsallis, C. 
\textit{Introduction to Nonextensive Statistical Mechanics: Approaching a Complex World}
(Springer, New York, NY, 2009).

\bibitem{Anteneodo}
Anteneodo, C. 
Non-extensive random walks,
{\it Physica A} \textbf{358}, 289--298 (2005).

\bibitem{SNT08}
Schw\"ammle, V., Nobre, F. D., \& Tsallis, C. 
q-Gaussians in the porous-medium equation: stability and time evolution,
{\it Eur. J. Phys. B} \textbf{66}, 537--546 (2008).

\bibitem{Ro10}
Romanelli, A. 
Distribution of chirality in the quantum walk: Markovian process and entanglement,
{\it Phys. Rev. A} \textbf{81}, 062349 (2010).

\bibitem{NPR}
Navarrete-Benlloch, C., P\'{e}rez, A., \& Rold\'{a}n, E. 
Nonlinear optical Galton board,
{\it Phys. Rev. A} {\bf 75}, 062333 (2007).

\bibitem{RMM04}
Ribeiro, P., Milman, P., \& Mosseri, R. 
Aperiodic quantum random walks,
{\it Phys. Rev. Lett.} \textbf{93} 190503 (2004).

\bibitem{Rohde}
Rohde, P. P., Brennen, G. K., \& Gilchrist, A. G. 
Quantum walks with memory provided by recycled coins and a memory of the coin-flip history
{\it Phys. Rev. A} {\bf 87}, 052302 (2013).

\bibitem{Yariv}
Yariv, A.
{\it Optical Electronics in Modern Communications}
(Oxford University Press, Oxford, 1997). 

\bibitem{Ro09}
Romanelli, A. 
The Fibonacci quantum walk and its classical trace map,
{\it Physica A} {\bf 388}, 3985--3990 (2009).

\bibitem{Mc}
McGettrick, M. 
One Dimensional Quantum Walks with Memory,
{\it Quantum Inf. Comp.} {\bf 10}, 0509--0524  (2010).

\bibitem{Joye}
Joye, A. \& Merkli, M. 
Dynamical Localization of Quantum Walks in Random Environments,
{\it J. Stat. Phys.} {\bf 140}, 1--29 (2010).

\bibitem{SK}
Shikano, Y. \& Katsura, H. 
Localization and fractality in inhomogeneous quantum walks with self-duality,
{\it Phys. Rev. E} {\bf 82}, 031122 (2010); 
Notes on Inhomogeneous Quantum Walks,
{\it AIP Conf. Proc.} {\bf 1363}, 151--154 (2011).

\bibitem{Ro04}
Romanelli, A., et al.
Quantum random walk on the line as a Markovian process,
{\it Physica A} \textbf{338}, 395--405 (2004).

\bibitem{GG96}
Godoy S. \& Garc\'ia-Col\'in, L. S. 
From the quantum random walk to classical mesoscopic diffusion in crystalline solids,
{\it Phys. Rev. E} \textbf{53}, 5779--5785 (1996).

\bibitem{Gaeta}
Gaeta, G. 
Asymptotic symmetries in an optical lattice,
{\it Phys. Rev. A} \textbf{72}, 033419 (2005).

\bibitem{mori}
Mori, H. 
Transport, Collective Motion, and Brownian Motion,
{\it Prog. Theor. Phys.} {\bf 33}, 423--455 (1965);
A Continued-Fraction Representation of the Time-Correlation Functions,
{\it Prog. Theor. Phys.} {\bf 34}, 399--416 (1965).

\bibitem{Olver}
Olver, P. J. 
{\it Applications of Lie Groups to Differential
Equations} (Springer-Verlag, New York,  NY, 1986).
\end{thebibliography}
\end{document}